\documentclass[12pt,preprint]{aastex}
\usepackage{psfig,epsfig,amsfonts,amsmath,amssymb,graphicx,morefloats,appendix,tensor}
\usepackage{color}
\usepackage{natbib}
\begin{document}

\slugcomment{Accepted by ApJ: July 7, 2016}

\title{ALMA Observations of the Debris Disk of Solar Analogue $\tau$ Ceti}

\author{Meredith A. MacGregor\altaffilmark{1}, 
Samantha M. Lawler\altaffilmark{2}, 
David J. Wilner\altaffilmark{1}, 
Brenda C. Matthews\altaffilmark{2,3}, 
Grant M. Kennedy\altaffilmark{4}, 
Mark Booth\altaffilmark{5}, 
James Di Francesco\altaffilmark{3,2}}
\altaffiltext{1}{Harvard-Smithsonian Center for Astrophysics, 60 Garden Street, Cambridge, MA 02138, USA}
\altaffiltext{2}{National Research Council of Canada, Herzberg Astronomy \& Astrophysics Program, Victoria, BC V9E 2E7, Canada }
\altaffiltext{3}{Dept. of Physics \& Astronomy, University of Victoria, Victoria, BC V8W 2Y2 Canada}
\altaffiltext{4}{Institute of Astronomy, Cambridge University, Cambridge CB3 0HA, United Kingdom}
\altaffiltext{5}{Instituto de Astrof\'{i}sica, Pontificia Universidad Cat\'{o}lica de Chile, 7820436 Macul, Santiago, Chile}

\begin{abstract}

We present 1.3~mm observations of the Sun-like star $\tau$ Ceti with the 
Atacama Large Millimeter/submillimeter Array (ALMA) that 
probe angular scales of $\sim1\arcsec$ (4 AU).  This first interferometric 
image of the $\tau$ Ceti system, which hosts both a debris disk 
and possible multiplanet system, shows emission from a nearly face-on belt of 
cold dust with a position angle of $90\degr$ surrounding an unresolved central source
at the stellar position.  To characterize this emission structure, we fit parametric
models to the millimeter visibilities.  The resulting best-fit model
yields an inner belt edge of $6.2^{+9.8}_{-4.6}$~AU, consistent with 
inferences from lower resolution, far-infrared \emph{Herschel} observations.  
While the limited data at sufficiently short baselines preclude us from 
placing stronger constraints on the belt properties and its relation to the proposed 
five planet system, the observations do provide a strong lower limit on the 
fractional width of the belt, $\Delta R/R > 0.75$ with $99\%$ confidence.
This fractional width is more similar to broad disks such as HD 107146 than 
narrow belts such as the Kuiper Belt and Fomalhaut.  
The unresolved central source has a higher flux density than the predicted flux
of the stellar photosphere at 1.3~mm.  Given previous measurements of an 
excess by a factor of $\sim2$ at 8.7~mm, this emission is likely due to a hot stellar chromosphere.

\end{abstract}


\section{Introduction}
\label{sec:intro}

The 5.8 Gyr-old \citep{mam08} 
main-sequence G8.5V star $\tau$ Ceti is the second closest \cite[3.65~pc,][]{vanL07} Solar-type star
reported to harbor both a tentative planetary system and a debris disk 
\cite[after $\epsilon$ Eridani, e.g.][]{gre98,hat00}.  The $\tau$ Ceti debris disk was 
first identified as an infrared excess by IRAS \citep{aum85} and 
confirmed by ISO \citep{hab01}.  \cite{gre04} marginally resolved 850 $\mu$m 
emission from the system with the James Clerk Maxwell Telescope (JCMT)/SCUBA, 
revealing a massive (1.2 $M_\oplus$) disk extending to 55 AU from the star.  Recent \emph{Herschel} 
observations at 70, 160, and 250~$\mu$m resolve the disk well and are best fit 
by a broad dust belt with an inner edge between $1-10$ AU and an outer edge at 
$\sim55$~AU \citep{law14}.  Due to its proximity and similarity to our Sun in age and spectral type, $\tau$ Ceti has been the 
object of numerous searches for planets using the radial velocity technique 
\cite[e.g.][]{pepe11}, most of which have proved unsuccessful.   
Using extensive modeling and Bayesian analysis of radial velocity data from the High Accuracy Radial Velocity Planet Searcher (HARPS) spectrograph \citep{may03,pepe11}, the Anglo-Australian Planet Search (AAPS) on the Anglo Australian Telescope (AAT),
and the High Resolution Echelle Spectrograph (HIRES) on the Keck telescope \citep{vogt94}, \cite{tuo13} report evidence for a tightly-packed five planet system.
This purported planetary system consists of five super-Earths
with masses of $4.0-13.2$~$M_\oplus$ (for orbits co-planar with the disk), semi-major axes 
ranging over $0.105-1.35$ AU, and small eccentricities, $e\sim0-0.2$.
The veracity of these planet candidates, however, remains controversial.  \cite{tuo13} acknowledge
that the detected signals could also result from a combination of instrumental bias
and stellar activity, although no further evidence is given to support these alternative
interpretations.  Also of note is the sub-Solar metallicity of $\tau$ Ceti, [Fe/H] $= -0.55\pm0.05$ dex \citep{pav12}, which makes it an interesting target for exoplanet searches due to the observed higher frequency of low-mass planets around low-metallicity stars \citep{jen13}.

We present interferometric observations of the $\tau$ Ceti system at 
1.3~mm using the Atacama Large Millimeter/submillimeter Array (ALMA).  
Millimeter imaging of this debris disk opens a unique window on the location 
and morphology of the underlying population of dust-producing planetesimals 
orbiting the star.  While these large, kilometer-sized bodies cannot be 
detected directly, millimeter observations probe emission from the large dust 
grains produced through collisions that are not rapidly redistributed by stellar radiation and winds \citep{wya06}.  
These new ALMA observations provide limits on the disk location and width,
which bear on the proposed planetary system within the disk. 
In Section~\ref{sec:obs}, we present the ALMA observations of the $\tau$ Ceti 
system.  In Section~\ref{sec:results}, we describe the analysis technique
and disk model results.  In Section~\ref{sec:disc}, we discuss the significance of the best-fit model parameters for 
the dust belt inner edge, width, proposed planetary 
system, and the origin of a bright, unresolved central emission source.

\section{Observations}
\label{sec:obs}

The $\tau$ Ceti system was observed using Band 6 (1.3~mm) in December 2014 with the ALMA 12-m array.  We obtained one scheduling block (SB) in good weather (PWV = 1.76~mm) with 34 antennas, with the longest baselines sampling to $1\arcsec$ ($4$ AU) resolution.
These observations were complemented by two SBs taken with the Atacama Compact 
Array (ACA) in July 2014 to provide shorter baselines and sensitivity to 
emission at larger scales.  For these ACA SBs, 11 operational antennas were 
available.  The observation dates, baseline lengths, and total time on-source are summarized in Table~\ref{tab:obs}.
For maximum continuum sensitivity, the correlator was configured to process two
polarizations in four 2 GHz-wide basebands 
centered at 226, 228, 242, and 244~GHz, each with 256 spectral channels. 
For the July SBs, the phase center was 
$\alpha = 01^\text{h}44^\text{m}02.348$, 
$\delta = -15\degr56\arcmin02\farcs509$ (J2000, ICRS reference frame).  The phase center for the December SB
was $\alpha = 01^\text{h}44^\text{m}02.299$, 
$\delta = -15\degr56\arcmin02\farcs154$ (J2000, ICRS reference frame).  Both phase centers were chosen to be the 
position of $\tau$ Ceti at the time of the observations given its proper motion of ($-1721.05$, $854.16$) 
mas yr$^{-1}$ \citep{vanL07}.  
The field of view is $\sim26\arcsec$, given by the FWHM size of the primary beam the ALMA 12-m antennas at the mean frequency of 234 GHz.

The data from all three SBs were calibrated separately using the \texttt{CASA} 
software package (version 4.2.2).  We corrected for time-dependent complex gain variations using 
interleaved observations of the calibrator J0132-1654.  Observations of J0137-2430 were used to determine the spectral response 
of the system. The absolute flux calibration scale was derived from observations of Neptune, and a mean calibration 
was applied to all four basebands, with a systematic uncertainty of $\sim10\%$ \cite[see][for a complete discussion of flux density models of Solar System bodies]{but12}. 

To generate a first image at the mean frequency, 234~GHz (1.3~mm), we Fourier inverted the calibrated visibilities with natural weighting and a multi-frequency synthesis with the \texttt{CLEAN} algorithm.  To improve surface brightness sensitivity, we included a modest taper using the \texttt{uvtaper} parameter in \texttt{CLEAN}, which controls the radial weighting of visibilities in the $(u,v)$-plane through the multiplication of the visibilities by the Fourier transform of a circular Gaussian (on-sky FWHM $= 6\arcsec$).  With the added taper, however, it became difficult to resolve the outer disk and the central stellar emission.  For clarity, we chose to image the disk and the star separately.  We isolate the disk emission by subtracting a point source model from these data using the CASA task \texttt{uvsub} to account for the stellar emission.  To isolate the stellar component, we image with \texttt{CLEAN} and no taper, only including baselines longer than 40 k$\lambda$, where we expect the star to dominate the emission (see Section~\ref{sec:results}). We choose to account for the primary beam in our modeling (see Section~\ref{subsec:modeling}) and thus do not apply a primary beam correction to any of these images.

\section{Results and Analysis}
\label{sec:results}

\subsection{Continuum Emission}
\label{subsec:continuum}

\begin{figure}[ht]
\begin{minipage}[h]{0.37\textwidth}
  \begin{center}
       \includegraphics[scale=0.5]{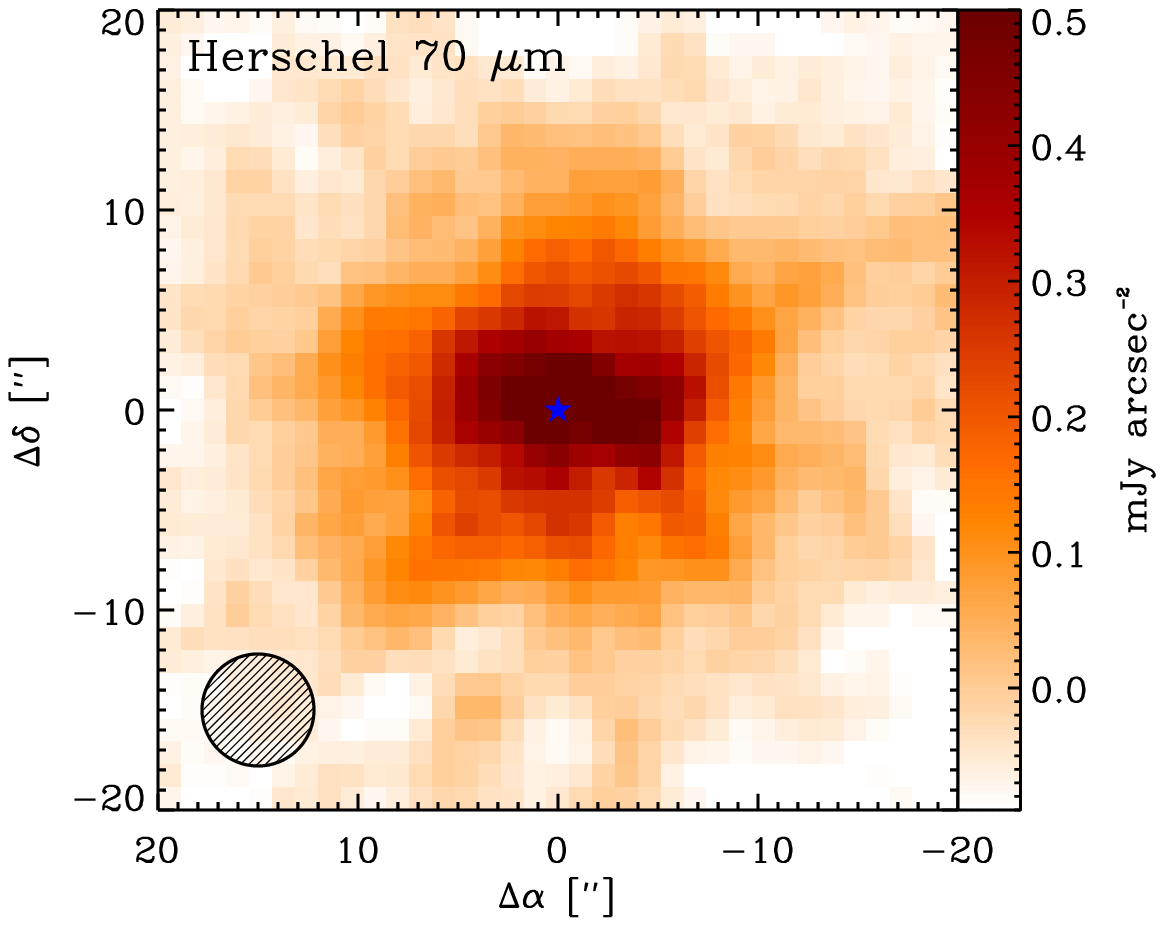}
  \end{center}
 \end{minipage}
 \begin{minipage}[h]{0.3\textwidth}
  \begin{center}
       \includegraphics[scale=0.5]{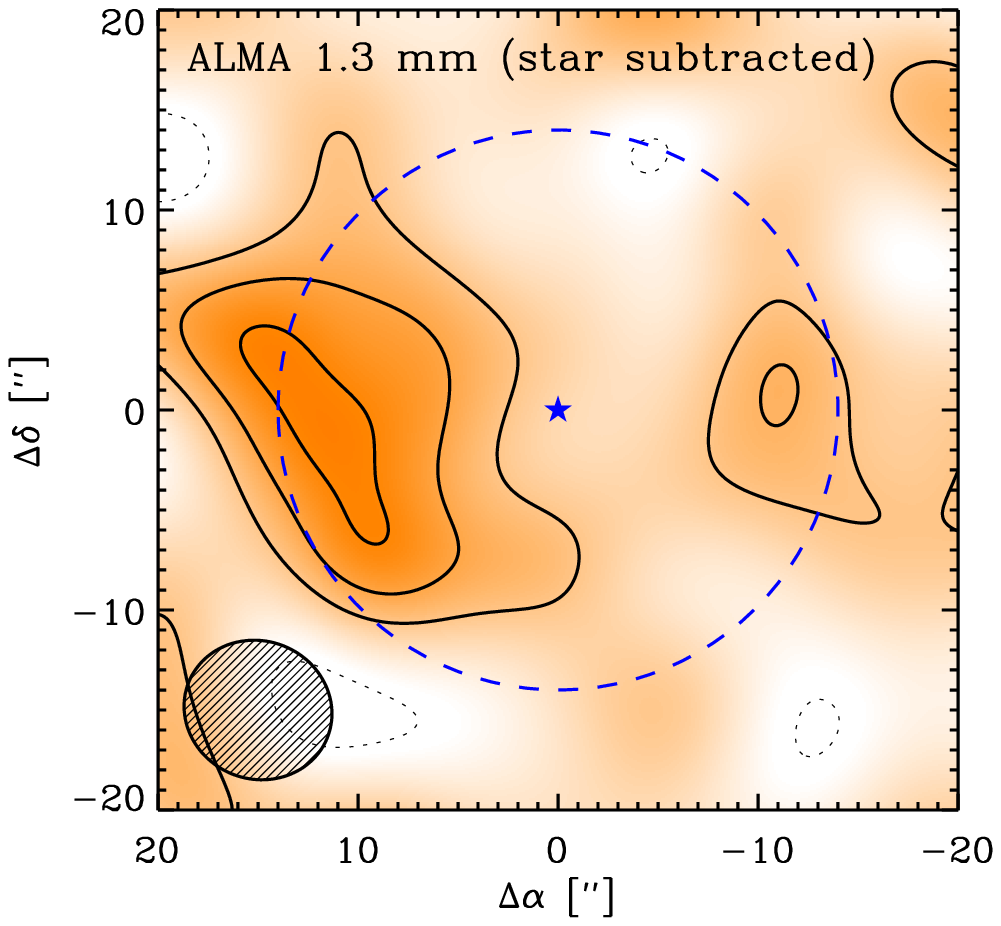}
  \end{center}
 \end{minipage}
  \begin{minipage}[h]{0.3\textwidth}
  \begin{center}
       \includegraphics[scale=0.5]{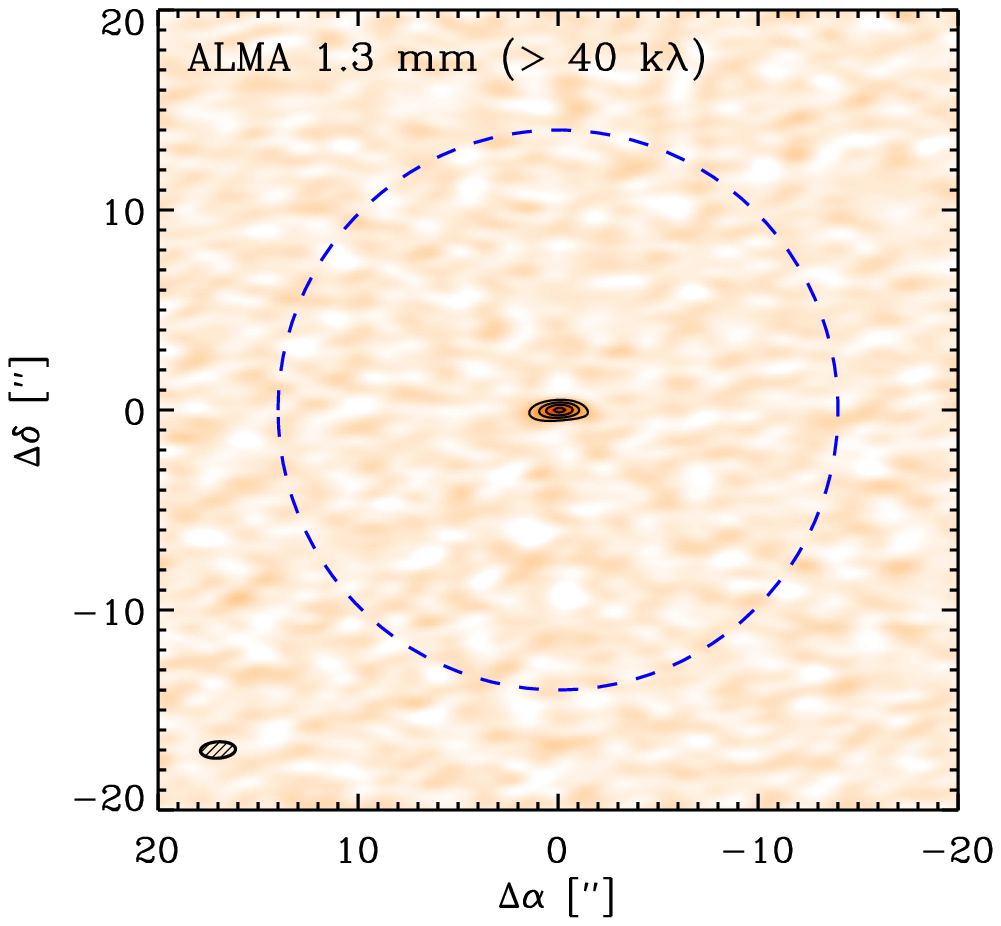}
    \end{center}
  \end{minipage}
\caption{\small \emph{(left)} \emph{Herschel/PACS} map of the 70~$\mu$m emission from the $\tau$ Ceti debris disk with the stellar contribution subtracted \cite[see][]{law14}.  The \emph{Herschel} $5\farcs6$ beam size is shown by the ellipse in the lower left corner.
\emph{(center)} The $\tau$ Ceti debris disk imaged by ALMA at 1.3~mm with contours in steps of $2\sigma$, where $\sigma$ is the rms noise level in the image $\sim30$~$\mu$Jy.  To isolate the disk emission, a point source model has been subtracted to account for the central stellar emission.  Using natural weighting along with a $6\arcsec$ Gaussian taper, the resulting FWHM synthesized beam size is $6\farcs5\times6\farcs1$.
\emph{(right)} ALMA image of the 1.3~mm continuum emission for baselines longer than 40~k$\lambda$ showing only the central point source with contours in steps of $5\sigma$.  Imaging with natural weighting and no taper yields a FWHM synthesized beam size of $1\farcs9\times1\farcs0$.
The position of the stellar photosphere is indicated in the left two panels by the blue star symbol.  The primary beam of the ALMA antennas at 1.3~mm (FWHM $\sim26\arcsec$) is shown by the dashed blue circle in the right two panels.
}
\label{fig:images}
\end{figure}

Figure~\ref{fig:images} shows an ALMA 1.3 mm image of the $\tau$ Ceti disk made 
with the central star subtracted (middle panel) along with an image including 
only baselines longer than 40 k$\lambda$ showing emission from the star and 
not the disk (right panel).  The \emph{Herschel/PACS} 70 $\mu$m star-subtracted
image (left panel) is shown for reference \citep{law14}. 
The natural weight rms noise is 30 $\mu$Jy and 180 $\mu$Jy for the 12-m and ACA observations, respectively.
For the image showing only the stellar emission, the natural weight rms is 
higher, 35 $\mu$Jy, since we exclude some baselines.
The belt is not detected in the ACA observations given the low signal-to-noise ratio, and
we only consider the 12-m data for imaging and modeling (see Section~\ref{subsec:modeling}).
For the 1.3~mm image of the star, the synthesized beam with natural weighting 
is $1\farcs9\times1\farcs0$ ($7\times4$~AU), and position angle $= -87\degr$.  
To improve surface brightness sensitivity, the image of the disk makes use of 
a modest taper and has a synthesized beam size of $6\farcs5\times6\farcs1$ 
($24\times22$~AU), and position angle $= 55\degr$.

These 1.3~mm images reveal (1) patchy emission ($\sim6\sigma$) from a nearly face-on (low inclination) dust disk,
and (2) a bright ($23\sigma$), unresolved central peak coincident with the expected stellar position.  The disk is located $\sim12\arcsec$ ($\sim44$~AU) from the star with a position angle of $\sim90\degr$ (E of N).  \cite{rei88} quantify the position uncertainty, $\sigma$ of a point source given the signal-to-noise ratio, $S/N$, and the synthesized beam size, $\theta$: $\sigma \sim 0.5\theta/(S/N)\approx 0\farcs14$, for our observations.  The position of the observed central source is coincident with the expected stellar position within this uncertainty.

\subsection{Emission Modeling Procedure}
\label{subsec:modeling}

We make use of the modeling scheme described in \cite{mac13,mac15b}.  In this approach, we construct parametric models of the 1.3~mm disk emission and then compute corresponding model visibilities using a python implementation\footnote{The code used to perform this part of the analysis is open source and freely available at \texttt{https:$//$github.com$/$AstroChem$/$vis\_sample}.} of the Miriad \texttt{uvmodel} task (Loomis et al. in prep).  To determine the best-fit parameter values and their uncertainties, we employ the \texttt{emcee} Markov Chain Monte Carlo (MCMC) package \citep{for13}.  This affine-invariant ensemble sampler for MCMC, enables us to accurately sample the posterior probability functions of all model parameters with minimal fine-tuning.  Due to the much higher rms noise of the ACA data, we choose to only fit models to the visibilities from the full 12-m ALMA array.

We model the millimeter emission of the $\tau$ Ceti debris disk as an axisymmetric, geometrically thin belt with an inner radius, $R_\text{in}$, an outer radius, $R_\text{out}$, and a radial surface brightness distribution described by a simple power law, $I_\nu \propto r^{\gamma-0.5}$.
Here, $\gamma$ describes the power law in radial surface density, 
$\Sigma \propto r^\gamma$, and temperature is assumed to follow a power law, 
$T\propto r^{-0.5}$, approximating radiative equilibrium for blackbody grains.  To first order, the dust temperature also depends on the grain opacity, $T\propto r^{-2/(4+\beta)}$, where $\beta$ is the power law index of the grain opacity as a function of frequency, $\kappa_\nu\propto\nu^\beta$.  \cite{gas12} measure $\beta = 0.58$, from observations of debris disks, which implies a temperature power law index of $\sim-0.44$.  Thus, the expected change in the temperature profile due to $\beta$ is much smaller than the uncertainty in our resulting model fits and we choose to ignore this effect.  Furthermore, the surface density and temperature profiles are degenerate, so we assume a blackbody profile and fit only for $\gamma$.

We constrain the outer disk radius using previous JCMT/SCUBA observations \citep{gre04}, since the parent body disk may have a different size relative to the smaller grains imaged with \emph{Herschel}.  While \cite{gre04} suggested that the disk was highly inclined, the \emph{Herschel} image (Figure~\ref{fig:images}, left panel) indicates that it is closer to face-on.  The SCUBA image is therefore marginally resolved at best, so we take their derived disk radius of 55~AU as an upper limit on $R_\text{out}$ and allow the inner radius, $R_\text{in}$, to vary.  We fit for the surface density radial power law index, $\gamma$, within a range of $-4$ to $4$.  The unresolved central peak seen in images is modeled by a central point source with flux, $F_\text{cen}$.  We do not fit for any relative offsets of the belt center, point source position, and phase center.
Models of the \emph{Herschel} images derive an inclination of $i = 35\degr\pm10\degr$ and position angle, $PA = 105\degr\pm10\degr$ \citep{law14}, and we assume that the millimeter belt emission is described by the same geometry.  For all parameters, we assume uniform priors and require that the model be physically plausible: $F_\text{cen} \geq 0$, and $0 \leq R_\text{in} < R_\text{out} \leq 55$~AU.

A total flux density, $F_\text{belt} = \int I_\nu d\Omega$, provides the normalization for the belt emission.  Using SCUBA on the JCMT, \cite{gre04} obtain a total flux density at 850 $\mu$m for the disk of $5.8\pm0.6$~mJy, including both the central star and likely contamination from background sources.  Recent SCUBA-2 observations at 850 $\mu$m yield a total flux density of $4.5\pm0.9$~mJy, including a contribution from the star of $\sim1$ mJy (Holland et al., in prep.).  An extrapolation of this measurement using the typical spectral index of 2.58 for debris disks at (sub)millimeter wavelengths \citep{gas12}, yields an expected flux density of the disk at 1.3 mm of $1.2\pm0.2$~mJy.  This more robust single-dish flux measurement allows us to constrain the total flux density of our models with a Gaussian prior, $0.6\text{ mJy}\leq F_\text{belt}\leq1.6\text{ mJy}$, accounting for uncertainty in both the single-dish 850~$\mu$m flux measurement and the extrapolation to 1.3~mm.

The angular scale of the $\tau$ Ceti debris disk is $\sim25\arcsec$ in diameter.  At 1.3~mm, the half power field of view of the 12-m ALMA antennas is comparable, FWHM$\sim26\arcsec$.  Given this, we must account for the effect of the primary beam response on our model parameters.  To do this, we model the ALMA primary beam as a Gaussian normalized to unity at the beam center and multiply each parametric disk model by this Gaussian beam model.  Since we account for the effect of the primary beam in our modeling scheme, we choose not to apply a primary beam correction to the images shown in Figure~\ref{fig:images} (right panels).

\subsection{Results of Model Fits}
\label{subsec:model_fits}

\begin{figure}[ht]
\centerline{\psfig{file=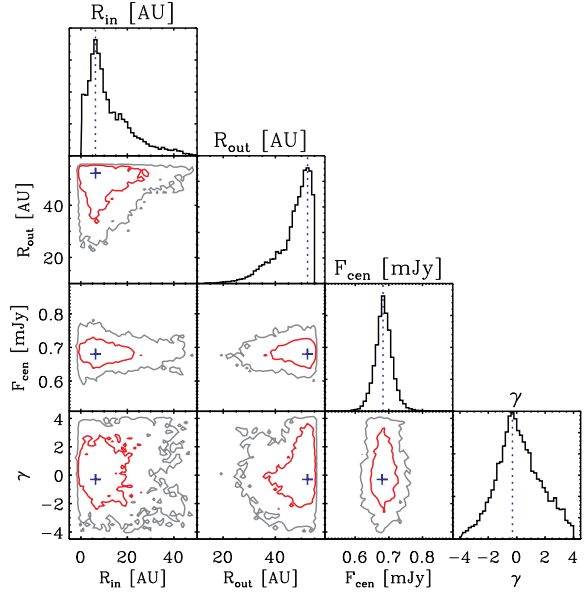,width=10cm,angle=0}}
\caption[]{\small The 1D (diagonal panels) and 2D (off-diagonal panels) projections of the posterior probability distributions for the best-fit model parameters ($R_\text{in}$, $R_\text{out}$, $F_\text{cen}$, and $\gamma$) resulting from $\sim10^4$ MCMC trials.  For a given parameter, the 1D distribution is determined by marginalizing over all other model parameters.  The best-fit parameter value is indicated by the vertical blue dashed line.  The 2D joint probability distributions show the $1\sigma$ (red) and $2\sigma$ (gray) regions for all parameter pairs, with the best-fit parameter values marked by the blue cross symbol.
}
\label{fig:mcmc}
\end{figure}

\begin{figure}[ht]
\centerline{\psfig{file=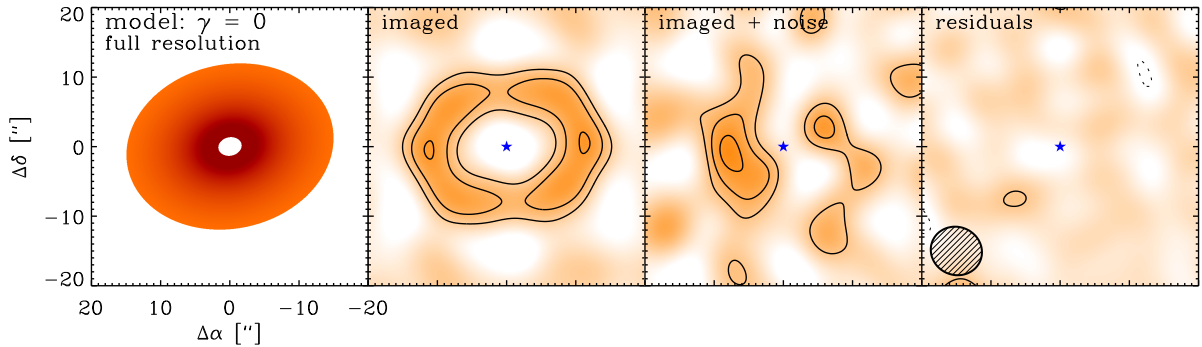,width=16cm,angle=0}}
\caption[]{\small  \emph{(left)} A full resolution (pixel scale $\sim0\farcs05\sim0.2$~AU) image of the best-fit model to the 1.3~mm ALMA continuum emission.  For simplicity, we have chosen a flat surface density profile with $\gamma=0$ and excluded the central stellar component.  \emph{(center left)} The same best-fit model convolved with the $\sim6\arcsec$ ALMA synthesized beam and imaged as in Figure~\ref{fig:images}, but with no noise added. \emph{(center right)}  The convolved best-fit model (same as shown in center left) with added simulated random noise at the same level as the ALMA 1.3~mm image, rms $\sim30$~$\mu$Jy. \emph{(right)} The residuals of the full best-fit model including the star and imaged with the same parameters as in Figure~\ref{fig:images}.  The ellipse in the lower left corner shows the $6\farcs5\times6\farcs1$ (FWHM) synthesized beam size.
}
\label{fig:model}
\end{figure}

Modeling the ALMA 1.3~mm visibilities yields a final best-fit model with a reduced $\chi^2$ value of 1.1.  Table~\ref{tab:mcmc} lists the best-fit values for each of the 5 free parameters along with their corresponding $1\sigma$ ($68\%$) uncertainties.  The 1D (diagonal panels) and 2D (off-diagonal panels) projections of the posterior probability distributions for all parameters except the total belt flux, $F_\text{belt}$, are shown in Figure~\ref{fig:mcmc}.  A full resolution image of this best-fit model (with a flat surface density profile, $\gamma=0$, and the central star excluded) is shown in the leftmost panel of Figure~\ref{fig:model}.  The same model convolved with the $\sim6\arcsec$ ALMA synthesized beam and imaged like the observations in Figure~\ref{fig:images} is shown in the next two panels both without (left) and with (right) simulated random noise with an rms of 30~$\mu$Jy.  Including the simulated noise results in a patchy image with emission structure similar to the ALMA 1.3~mm image shown in Figure~\ref{fig:images}.  In both the ALMA and model images, the most significant peaks of emission are consistent with the expectation for a slightly inclined disk with PA near $90\degr$.  The rightmost panel of Figure~\ref{fig:model} shows the residuals resulting from subtracting this best-fit model from the observed visibilities, again imaged with the same parameters.  No significant features are evident.

The best-fit total belt flux density is $F_\text{belt} = 1.0^{+0.6}_{-0.4}$~mJy, constrained by the Gaussian prior taken from previous single dish flux measurements.  \cite{law14} note that the SCUBA and SCUBA-2 flux densities are higher than expected given an extrapolation from the \emph{Herschel} flux density measurements.  This difference suggests that these earlier observations could be contaminated by the extragalactic background or that the disk could have an additional warm component.  Given the limits in sensitivity of our ALMA data, the total flux density we measure is consistent with both the \emph{Herschel} and SCUBA/SCUBA-2 values and we cannot distinguish between these two scenarios.

Not surprisingly, given the sensitivity limits of the ALMA data, model fitting does not provide a strong constraint on the power law index of the surface density radial profile, $\gamma=-0.3^{+1.9}_{-1.3}$.  With large uncertainty, this result implies a shallow surface density profile.  In addition, we see a clear degeneracy between the surface density gradient, $\gamma$, and the disk outer radius, $R_\text{out}$ \cite[e.g.][]{mun96}.  For very negative values of $\gamma$, the outer regions of the resulting belt model have low surface brightness, making it difficult to constrain the position of the outer edge.  Thus, the contours shown in Figure~\ref{fig:mcmc} for that pair of parameters exhibit a slope, spreading out to span a wide range of possible outer radii for increasingly negative values of $\gamma$.

\begin{figure}[ht]
\centerline{\psfig{file=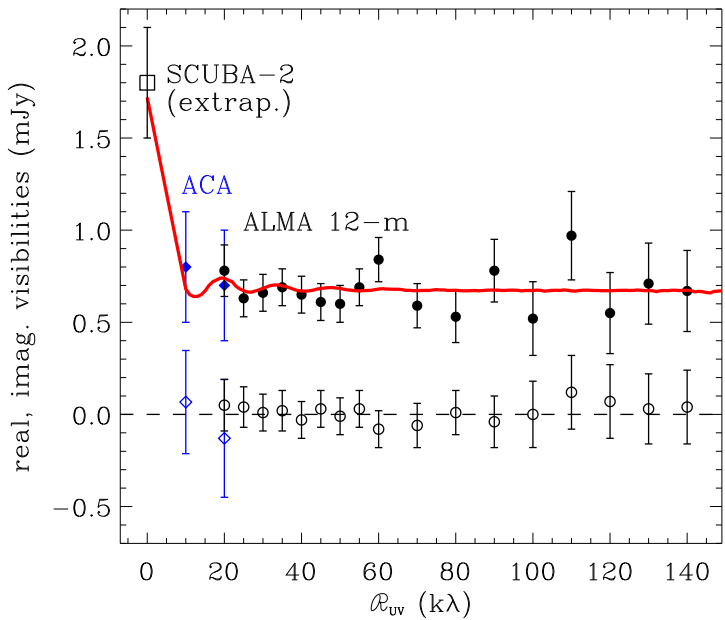,width=10cm,angle=0}}
\caption[]{\small The deprojected real (filled symbols) and imaginary (open symbols) visibilities for the ACA (blue diamonds) and 12-m array (black circles), compared to the best-fit belt model (red solid line).  The single dish SCUBA-2 flux (Holland et al., in prep.) extrapolated from 850 $\mu$m to 1.3 mm is also plotted at $\mathcal{R}_{uv} = 0$ k$\lambda$.
}
\label{fig:vis}
\end{figure}

Another helpful way to visualize and compare the ALMA observations and the best-fit model is by deprojecting the real and imaginary visibilities based on the inclination, $i$, and position angles, $PA$, of the disk major axis, as is shown in Figure~\ref{fig:vis} \cite[see][for a detailed description of deprojection]{lay97}.  Essentially, the coordinates for each visibility point are defined by a distance from the origin of the $(u,v)$ plane, $\mathcal{R} = \sqrt{u^2+v^2}$. To change to a deprojected, rotated coordinate system, we define an angle $\phi = \frac{\pi}{2}-PA$, where $PA$ is the position angle of the disk measured east of north.  The new coordinates are defined as $u' = u\text{ cos}\phi+v\text{ sin}\phi$ and $v' = (-u\text{ sin}\phi+v\text{ cos}\phi)\text{ cos}i$, where $i$ is the inclination angle of the disk.  Then, the new deprojected $(u,v)$ distance is $\mathcal{R}_{uv} = \sqrt{u'^2+v'^2}$.  Assuming that the disk is axisymmetric, we average the visibilities azimuthally in annuli of $\mathcal{R}_{uv}$.  For our ALMA $\tau$ Ceti observations, the real part of the deprojected visibilities is reasonably consistent with the prediction for a broad belt of emission, showing a central peak and several oscillations of decreasing amplitude.  The constant offset from zero is the visibility signature of the unresolved central peak we see clearly in the images.  The imaginary visibilities are essentially zero, indicating that there is no asymmetric structure in the disk, which is consistent with the absence of any significant residuals in Figure~\ref{fig:model} (rightmost panel).  Note that we are lacking $(u,v)$ coverage on baselines shorter than $\lesssim20$ k$\lambda$, the region of the visibility curve with the most structure.

\section{Discussion}
\label{sec:disc}

We have obtained ALMA 1.3~mm observations of the $\tau$ Ceti system using both the ACA and the full 12-m array with baselines corresponding to scales of $1\arcsec$ ($4$ AU). The resulting image shows emission from an outer dust disk located $\sim12\arcsec$ ($\sim44$ AU) from the star surrounding an unresolved central peak.  We fit parametric models to the millimeter visibilities, which included two components: (1) an outer disk with a radial surface density profile described by a power law with index $\gamma$, and (2) a point source at the stellar position.  In the context of our simple model, this analysis provides tentative constraints on the location of the disk inner edge and the width of the disk.  We now compare the model fits to previous \emph{Herschel} observations and discuss implications for the geometry of the proposed inner planetary system located within the dust belt.

\subsection{Location of the Disk Inner Edge and Belt Width}
\label{subsec:inner_edge}

Our best-fit model yields an inner radius for the disk of $6.2^{+9.8}_{-4.6}$~AU, consistent with the analysis of \emph{Herschel} observations that constrained the inner edge of the disk to be between 1 and 10~AU from the star \citep{law14}.  For comparison, the planetary system proposed by \cite{tuo13} consists of five super-Earths in a tightly-packed configuration with semi-major axes ranging over $0.105-1.35$~AU.  Given the uncertainties on $R_\text{in}$ from our best-fit model, the disk could extend well into this inner planetary system ($R_\text{in} < 1$ AU) or end far beyond the outermost planet ($R_\text{in} > 2$ AU).  None of the proposed planets have large enough orbital radius or mass to cause significant perturbations or clear the disk beyond 3 AU (within the range of $R_\text{in}$ allowed by our models). Lawler et al. (2014) use numerical simulations to show that the system would be stable with an additional Neptune-mass planet on an orbit of $5-10$ AU, the largest mass planet at such separations that cannot be ruled out by the radial velocity data.

\begin{figure}[ht]
\centerline{\psfig{file=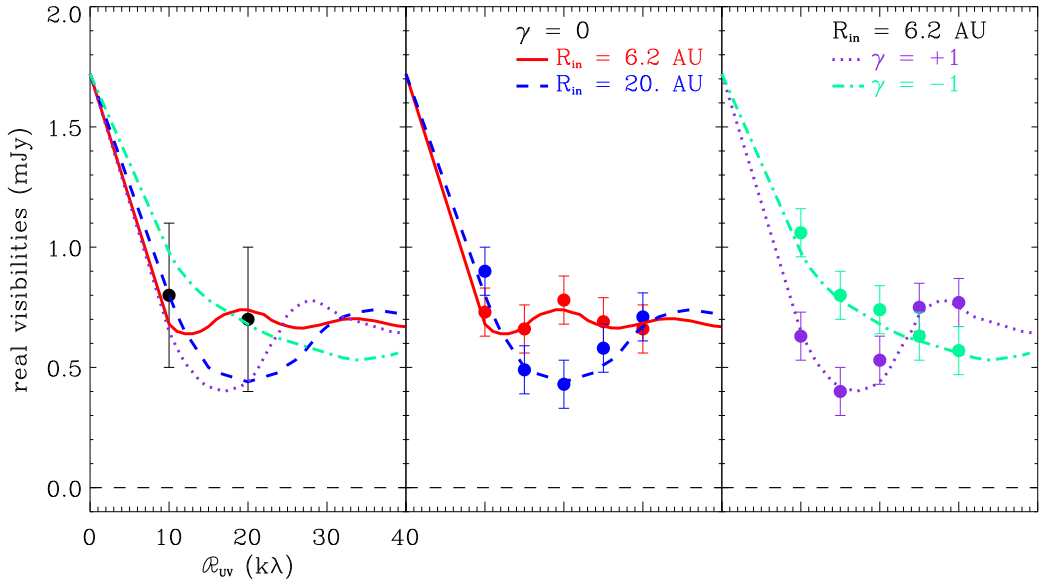,width=14cm,angle=0}}
\caption[]{\small \emph{(left)}~The deprojected real component of the expected complex visibilities for belt models with our best-fit $R_\text{in}=6.2$~AU and $\gamma=-1,0,+1$ (dot-dash green line, solid red line, and dotted purple line, respectively), and a model with $R_\text{in} = 20$~AU and $\gamma=0$ (dashed blue line).  The real visibilities from our ACA observations presented here are shown by the black points and are consistent with all four models. 
\emph{(center)}~The real visibilities of simulated ACA 1.3~mm emission for models with $\gamma=0$ and $R_\text{in}=6.2$ and $20$~AU (red and blue points, respectively).  With 10 antennas and 10 hours on source, these models are easily distinguishable. 
\emph{(right)}~The real visibilities of simulated ACA 1.3~mm emission for models with $R_\text{in}=6.2$~AU and $\gamma=+1$ and $-1$ (purple and green points, respectively).  Again, these profiles are clearly different in shape, with the zero-crossing null locations shifted by $>10$ k$\lambda$.
}
\label{fig:sim}
\end{figure} 

The belt position and width are strongly constrained by the location of the first null in the deprojected real visibilities \cite[see Figure~\ref{fig:vis},][]{mac15b}.  Although we obtained some ACA data, the integration time was short, and the resulting sensitivity (rms $\sim180$~$\mu$Jy) at short baselines ($< 20$~k$\lambda$) was insufficient to discriminate between disk models with inner radii of $1-10$~AU, the parameter space with significant implications for the proposed planetary system.  New observations with shorter baselines are needed to better determine the location of the dust belt, as well as its radial surface density gradient. To demonstrate the contribution that such observations would make to our analysis, we carried out simulations of ALMA ACA observations (rms 60~$\mu$Jy, using 10 antennas in the Cycle 4 setup) at 1.3~mm for a model with our best-fit $R_\text{in} = 6.2$ AU and $\gamma=-1,0,+1$,  and a model with $R_\text{in} = 20$ AU and $\gamma=0$, all consistent with the ALMA observations presented here. Figure~\ref{fig:sim} (left panel) shows the real component of the expected complex visibilities for all four models, along with our current ACA observations.  The center and right panels show the real part of simulated ACA visibilities for all four belt models compared to the expected theoretical visibility curves.  These profiles are clearly different in shape, with the zero-crossing locations shifted by $>10$ k$\lambda$ and the amplitude of the oscillations differing by more than a factor of 2.

Although the ALMA observations allow for broad disk models that extend in toward the 
central star, they are not consistent with a narrow ring model located far 
from the star.  The contours for the inner and outer radius in Figure~\ref{fig:mcmc} show the absence of any models with large $R_\text{in}$ and small $R_\text{out}$, indicating that the disk must be broad.  Indeed, we can place a strong upper limit, $R_\text{in}<25$ AU with $99\%$ ($3\sigma$) 
confidence.  Given the values of $R_\text{in}$ and $R_\text{out}$ from our best-fit model, the 
fractional width of the $\tau$ Ceti disk is $\Delta R/R = 1.6^{+0.3}_{-0.6}$. 
If we assume that the outer belt edge at millimeter wavelengths aligns 
with the edge found at far-infrared wavelengths ($R_\text{out} = 55$~AU), we can place
a lower limit on the belt width, $\Delta R > 30$~AU.  
 At $99\%$ confidence, $\Delta R/R > 0.75$.  
 For comparison, our Solar System's classical Kuiper Belt has a fractional width of $\Delta
R/R \sim 0.18$ \cite[e.g.][]{hah05,ban15}, significantly more narrow.  In fact, the Kuiper Belt
appears to be confined between Neptune's 3:2 and 2:1 resonances.  Similarly, the Fomalhaut
debris disk appears narrow with $\Delta R /R \sim0.1$, possibly attributable to planets orbiting both interior to
and exterior to the ring \citep{bol12}.  In contrast, recent ALMA observations of 
the HD~107146 debris disk \citep{ric15} indicate that its belt extends from 
30~AU to 150~AU ($\Delta R/R \sim 1.3$), with a break at $\sim70$AU.
The $\epsilon$ Eridani debris disk also appears to be somewhat broader with a
fractional width determined from resolved SMA observations of $\Delta R/R = 0.3$ \citep{mac15b}.  
The fractional width of the $\tau$ Ceti belt is 
substantially larger than both the classical Kuiper Belt and Fomalhaut.  
However, the $\tau$ Ceti belt is comparable in width to the HD 107146 disk, 
suggesting that it might also have a more complicated radial structure,
which we are unable to resolve with these observations.

\cite{kal06} discuss the implications of the observed diversity in debris disk structures
in the context of scattered light observations.  For a narrow belt structure, both
the inner and outer edges of the disk must be maintained by gravitational perturbers
such as stellar or substellar companions, or be confined by mean-motion resonances with an interior 
planet as is the case for our own Kuiper Belt.  Without any such confinement mechanism
for the outer disk edge, and since more massive planets have been ruled out around 
$\tau$ Ceti at distances approaching $\sim10$ AU \citep{law14}, the expected structure is indeed a wide belt.

\subsection{Central Component}
\label{subsec:star}

In addition to the extended emission from an outer belt, the ALMA 1.3~mm image shows a bright, unresolved point source (see the constant positive offset on long baselines in Figure~\ref{fig:vis}) at the expected position of the star with a flux density of $0.69^{+0.02}_{-0.05}$ mJy.  For a G8.5V star with an effective temperature of $5344\pm50$~K, an extrapolation of a PHOENIX stellar atmosphere model \citep{hus13} predicts a 1.3 mm flux density of 0.60~mJy (with 5\% uncertainty).  Thus, the flux density of this central source is marginally higher than the expectation for the stellar photosphere at this millimeter wavelength.  We note, however, that an extrapolation of the mid-infrared flux of the star, as measured by WISE  at 22~$\mu$m \citep{wri10} and AKARI at 9 and 18~$\mu$m \citep{ish10}, yields a prediction for the flux of the stellar photosphere at 1.3~mm of $\sim0.5$~mJy, substantially lower than the measured 1.3~mm flux density.  Our ALMA measurement is complemented by previous observations by \cite{vill14} with the Karl G. Jansky Very Large Array (VLA) at 34.5~GHz (8.7~mm) and 15.0~GHz (2.0~cm).  At 8.7~mm, the measured flux density is $25.3\pm3.9$~$\mu$Jy, significantly higher than the predicted photospheric flux density of 14~$\mu$Jy.  While the star is not detected at 2.0~cm, a robust 99\% confidence upper limit is determined of $<11.7$~$\mu$Jy (model photospheric prediction of 2.5~$\mu$Jy).

\begin{figure}[ht]
\centerline{\psfig{file=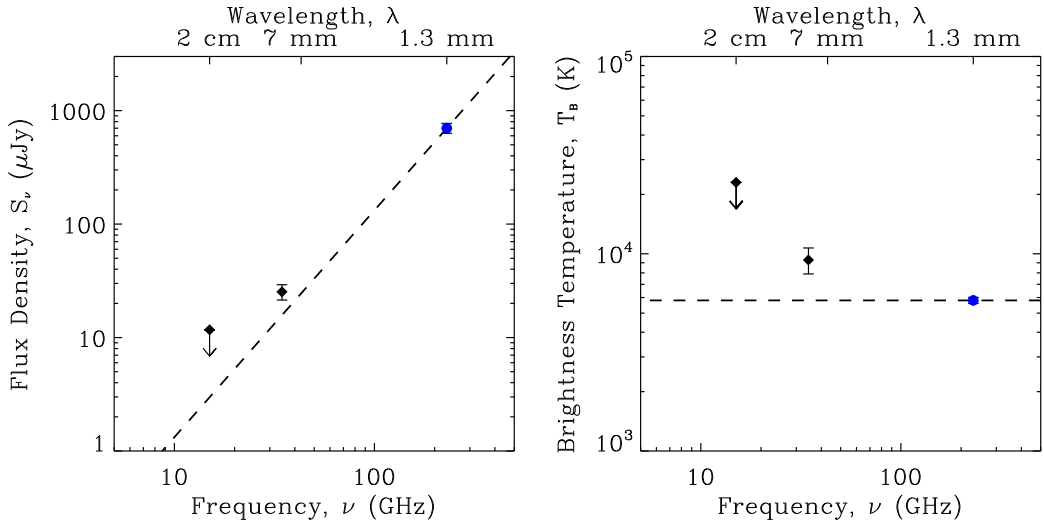,width=14cm,angle=0}}
\caption[]{\small \emph{(left)} Flux density spectrum of $\tau$ Ceti from ALMA and VLA observations. The dashed line indicates the expected spectral index of 2.0 for a classical photosphere. \emph{(right)} Brightness temperature spectrum calculated assuming the photospheric radius of the star. For both plots, our ALMA measurements are shown as blue circles and the VLA measurements \citep{vill14} are shown as black diamonds.  Detections are indicated by points with $1\sigma$ error bars.  The 99\% upper confidence limit at 2.0 cm is indicated by the downwards arrow.  Again, the dashed line indicates the expected brightness temperature for a classic photosphere with the brightness temperature determined from our 1.3~mm ALMA measurement.
}
\label{fig:star}
\end{figure}

As \cite{vill14} discuss, the observed unresolved emission from $\tau$ Ceti at both millimeter and centimeter wavelengths plausibly arises from a hot stellar chromosphere.  Similar excess emission at long wavelengths has been noted for several neighboring Sun-like stars, including $\alpha$ Cen A and B (spectral types G2V and K2V, respectively) observed with ALMA by \cite{lis15} and $\epsilon$ Eridani (spectral type K2V) observed with the Submillimeter Array (SMA) and Australia Telescope Compact Array (ATCA) by \cite{mac15b}.  We combine our new ALMA 1.3~mm flux density with the previous VLA 8.7~mm measurement and 2~cm upper limit, and determine the Planck brightness temperature at all three wavelengths \cite[following][]{lis13}.  Figure~\ref{fig:star} shows the resulting ALMA and VLA constraints on both the flux density and the brightness temperature spectra of $\tau$ Ceti. We assume that the photospheric radius is comparable at optical and millimeter/centimeter wavelengths, and adopt a value of $0.793\pm0.004$ $R_\odot$, obtained from interferometric measurements using the FLUOR instrument on the CHARA array \citep{dif07}.  At 1.3~mm this analysis yields $T_B = 5,800\pm200$ K, modestly hotter than the effective temperature of $5344\pm50$~K.  However, at longer wavelengths, the brightness temperature diverges significantly from the photospheric prediction with $T_B = 9,300\pm1400$~K and $<23,000$~K at 8.7~mm and 2~cm, respectively.

Additionally, the spectral index at long wavelengths of the central emission from $\tau$ Ceti shows the same deviation from an optically thick photosphere (spectral index of $\sim2$) as is seen for $\alpha$ Cen A and B and $\epsilon$ Eridani.  Between 1.3 and 8.7~mm, the spectral index of the central peak in our observations of $\tau$ Ceti is $1.74\pm0.15$ (with the $\sim10\%$ uncertainty in the flux scale and the $1\sigma$ modeling errors added in quadrature).  For comparison, the measured spectral indices between 0.87 and 3.2~mm are 1.62 and 1.61 for $\alpha$ Cen A and B, respectively \citep{lis15}.

\section{Conclusions}
\label{sec:conclusions}

We observed the $\tau$ Ceti debris disk with ALMA at 1.3~mm with baselines that probe $1\arcsec$ (4~AU) scales.  These are the first observations of this nearby system with a millimeter interferometer and reveal somewhat patchy emission from a dust disk surrounding an unresolved central emission peak.  In order to characterize these two emission components, we fit simple parametric models directly to the visibility data within an MCMC framework.

Our best-fit model yields an inner belt edge of $6.2^{+9.8}_{-4.6}$ AU, 
consistent with the analysis of previous far-infrared \emph{Herschel} 
observations.  Given the relatively low sensitivity at short baselines in 
the ALMA observations, we are unable to place a tighter constraint on the 
inner edge and its position relative to the proposed five planet system.  
These data, however, provide a strong lower limit on the fractional width of 
the belt, $\Delta R/R > 0.75$ with $99\%$ confidence.  This result implies that 
the $\tau$ Ceti debris disk is broad, much wider than the classical Kuiper Belt
in our Solar System and more comparable to the HD 107146 debris disk \citep{ric15}. 

The bright central peak at the stellar position has a flux density of $F_\text{1.3mm}=0.69^{+0.02}_{-0.05}$~mJy, somewhat higher than the predicted flux of the stellar photosphere at 1.3~mm.  At longer centimeter wavelengths, this excess is more significant, increasing to $\sim2\times$ the photospheric prediction in VLA observations at 8.7~mm \citep{vill14}.  The spectral index between these two measurements is $1.74\pm0.15$, shallower than the expectation for an optically thick photosphere.  Given the high brightness temperatures at both 1.3 and 8.7~mm, this excess emission is likely due to a hot stellar chromosphere. Similar spectra have been observed for other nearby Sun-like stars, e.g. $\alpha$ Cen A/B and $\epsilon$ Eridani.  

These first ALMA observations of the $\tau$ Ceti system allow us to probe the structure of the debris disk with higher resolution than previous work.  However, higher sensitivity observations at shorter baselines are still needed to constrain 
the location of the inner edge of the dust belt more precisely.  If the disk extends in towards the star, within the orbit of the outermost proposed planet, this provides strong evidence against the posited five planet system.
However, if the disk inner edge is located well outside the proposed planetary system, an additional massive planet on a wide orbit may be required to clear out the central hole in the belt.  Additional observations with the ACA could provide the necessary sensitivity to determine the position of the inner disk edge and its implications for an interior planetary system.

\acknowledgements
This paper makes use of the following ALMA data: ADS/JAO.ALMA\#2013.1.00588.S. 
ALMA is a partnership of ESO (representing its member states), NSF (USA) and 
NINS (Japan), together with NRC (Canada) and NSC and ASIAA (Taiwan) and KASI 
(Republic of Korea), in cooperation with the Republic of Chile. The Joint ALMA 
Observatory is operated by ESO, AUI/NRAO and NAOJ. The National Radio Astronomy
Observatory is a facility of the National Science Foundation operated under 
cooperative agreement by Associated Universities, Inc. M.A.M acknowledges 
support from a National Science Foundation Graduate Research Fellowship 
(DGE1144152).  S.M.L. gratefully acknowledges support from the NRC Canada Plaskett Fellowship.
B.C.M. acknowledges support from a Natural Science and 
Engineering Research Council (NSERC) Discovery Accelerator Supplement grant. 
G.M.K. is supported by the Royal Society as a Royal Society University Research Fellow. 
M.B. acknowledges support from a FONDECYT Postdoctral Fellowship, project no. 3140479 and the Millennium Science Initiative (Chilean Ministry of Economy), through grant RC130007.

\bibliography{References}
\pagebreak

\begin{deluxetable}{l c c c c}
\tablecolumns{5}
\tabcolsep0.06in\footnotesize
\tabletypesize{\small}
\tablewidth{0pt}
\tablecaption{ALMA Observations of $\tau$ Ceti}
\tablehead{
\colhead{Observation} & 
\colhead{Array} & 
\colhead{\# of } & 
\colhead{Projected} & 
\colhead{Time on} \\
\colhead{Date} & 
\colhead{} & 
\colhead{Antennas} & 
\colhead{Baselines (m)} &
\colhead{Target (min)}
}
\startdata
2014 Jul 7 & ACA & 11 & $9-50$ & 5.8\\
2014 Jul 16 & ACA & 11 & $9-50$ & 33.9  \\
2014 Dec 15 & 12-m & 34 & $15-350$ & 41.4 
\enddata
\label{tab:obs}
\end{deluxetable}

\begin{deluxetable}{l l c c}
\tablecolumns{4}
\tabcolsep0.1in\footnotesize
\tabletypesize{\small}
\tablewidth{0pt}
\tablecaption{ALMA Model Parameters}
\tablehead{
\colhead{Parameter} & 
\colhead{Description} & 
\colhead{Best-fit} &
\colhead{68\% Confidence Interval}
}
\startdata
$R_\text{in}$ & Belt inner radius (AU) & $6.2$ & $+9.8,-4.6$ \\
$R_\text{out}$ & Belt outer radius (AU) & $52.$ & $+3.,-8.$ \\
$F_\text{belt}$ & Belt flux density (mJy) & $1.0$ & $+0.6,-0.4$\\
$F_\text{cen}$ & Central source flux (mJy) & $0.69$ & $+0.02,-0.04$ \\
$\gamma$ & Belt surface density power law index & $-0.3$ & $+1.9,-1.3$\\
\enddata
\label{tab:mcmc}
\end{deluxetable}

\end{document}